\documentclass{aa}  % two-column style
\usepackage{epsfig}			% For eps figures, old commands
\usepackage{graphicx,color}		% For eps figures, newer & more powerfull
\usepackage{amssymb}			% For useful mathematical symbols
\usepackage{color}			% For color text: \color command
\usepackage{url}			% For breaking URLs easily trough lines
\usepackage{amsmath}			% Provides various features to facilitate writing math formulas
\usepackage{rotating}			% For rotate any object of an arbitrary angle
\usepackage{float}			% For improveing the interface for defining floating objects
\usepackage{textcomp}			% Pro­vides many text sym­bols (such as baht, bul­let, copy­right,etc )
\usepackage{epstopdf}
\usepackage{dcolumn}
\usepackage{times}
\usepackage{tabularx}
\usepackage{capt-of}
\usepackage[normalem]{ulem}
\begin{document}

%%%%%---------------Title----------------%%%%%

\title{Sunspot area catalogue revisited:\\ Daily cross-calibrated areas since 1874\thanks{Generated composites are available online at http://www2.mps.mpg.de/projects/sun-climate/data.html or at the CDS via anonymous ftp to cdsarc.u-strasbg.fr (130.79.128.5) or via http://cdsweb.u-strasbg.fr/cgi-bin/qcat?J/A+A/}}
\titlerunning{Sunspot area catalogue since 1874}
   \author{Sudip Mandal
          \inst{1}
          \and
          Natalie A. Krivova\inst{1}
          \and
          Sami K. Solanki\inst{1,2}
          \and
          Nimesh Sinha\inst{1}
          \and
          Dipankar Banerjee\inst{3,4,5}
          }

   \institute{Max Planck Institute for Solar System Research, Justus-von-Liebig-Weg 3, 37077, G{\"o}ttingen, Germany \\
              \email{smandal.solar@gmail.com}, ORCID: \url{0000-0002-7762-5629}
         \and
             School of Space Research, Kyung Hee University, Yongin, Gyeonggi 446-701, Republic of Korea
         \and
             Indian Institute of Astrophysics, Koramangala, Bangalore 560034, India
         \and
             Center of Excellence in Space Sciences India, IISER Kolkata, Mohanpur 741246, West Bengal, India
         \and
             Aryabhatta Research Institute of Observational Sciences, Nainital 263000, Uttarakhand, India
             }

\abstract
{Long and consistent sunspot area records are important for understanding the long-term solar activity and variability. Multiple observatories around the globe have regularly recorded sunspot areas, but such individual records only cover restricted periods of time. Furthermore, there are also systematic differences between them, so that these records need to be cross-calibrated before they can be reliably used for further studies.}
{We produce a cross-calibrated and homogeneous record of total daily sunspot areas, both projected and corrected, covering the period between 1874 and 2019. A catalogue of calibrated individual group areas is also generated for the same period.}
{We have compared the data from nine archives: Royal Greenwich Observatory (RGO), Kislovodsk, Pulkovo, Debrecen, Kodaikanal, Solar Optical Observing Network (SOON), Rome, Catania, and Yunnan Observatories, covering the period between 1874 and 2019. Mutual comparisons of the individual records have been employed to produce homogeneous and inter-calibrated records of daily projected and corrected areas. As in earlier studies, the basis of the composite is formed by the data from RGO. After 1976, the only datasets used are those from Kislovodsk, Pulkovo and Debrecen observatories.
This choice was made based on the temporal coverage and the quality of the data. While there are still 776 days missing in the final composite, these remaining gaps could not be filled with data from the other archives as the missing days lie either before 1922 or after 2016 and none of the additional archives covers these periods.
}
{In contrast to the SOON data used in previous area composites for the post-RGO period, the properties of the data from Kislovodsk and Pulkovo are very similar to those from the RGO series. They also directly overlap the RGO data in time, which makes their cross-calibration with RGO much more reliable. Indeed, comparing our area catalogue with previous such composites, we find improvements both in data quality and coverage. We have also computed and provide the daily Photometric Sunspot Index (PSI) widely used, e.g., in empirical reconstructions of solar irradiance.}
{}

   \maketitle
   
%=================INTRODUCTION======================
\section{Introduction}

Sunspots, the largest known dark photospheric features, are probably the most famous manifestation of solar activity. Solar activity is driven and modulated by a common process, the solar magnetic field and its interaction with solar convection. Sunspots are one of the oldest (although indirect) measurements of the solar magnetic fields. Hence, sunspot area records play an important role in our understanding of the long term behaviour of solar magnetic activity and variability.

 Barring few individual measurements (see \citealt{2007AdSpR..40..929V} for a review of historical sunspot observations), systematic monitoring of sunspot area started at the Royal Greenwich Observatory (RGO) in 1874. RGO recorded daily areas and positions of sunspots. In the 20th century, various observatories around the world (e.g. Kodaikanal, Pulkovo, Mt. Wilson, Kislovodsk, to name a few), also initiated  similar observing programs and started accumulating sunspot data. After continuing for a century, RGO stopped its campaign in 1976 and transferred the program to Debrecen observatory, where such area observations are still carried out on a daily basis. If all these available records are stitched together, the combined series covers a period of almost 150 years, which yields a data set suitable for studies of the long-term changes in solar magnetism.

Such a composite series is extremely important for multiple solar applications. For example, individual sunspot group areas are required for reconstructions of the long-term evolution of the solar surface magnetic field \citep[e.g.,][]{2011A&A...528A..83J,2014SSRv..186..491J}, estimates of the solar radiative flux suppression via the Photometric Sunpot Index (PSI; \citealp{1994SoPh..152..119B}), 
or assessment of the sunspot magnetic field and its long-term changes \citep{2014SoPh..289.1143T,2017AN....338...26N}, while historical solar irradiance reconstructions \citep[e.g.,][]{Foukal556,2000A&A...353..380F,2007A&A...467..335K,2010JGRA..11512112K,2014A&A...570A..23D,2016A&A...590A..63D,2017JGRA..122.3888Y} often also use the daily total areas as input.
Understanding and reconstructions of the past solar variability are, in turn, important for an assessment of the solar influence on Earth's climate (\citealp[see., e.g.,][]{2013ARA&A..51..311S}).

It is, therefore, not surprising that significant effort has been made towards cross-calibrating the various individual sunspot area datasets \citep{Nagovitsyn1997, 1997SoPh..173..427F,2001MNRAS.323..223B,2002SoPh..211..357H,2009JGRA..114.7104B,2013MNRAS.434.1713B}. This is, however, not a trivial task. Deviations in the observing facilities, seeing conditions, capturing devices, data processing techniques, etc., introduce partly significant systematic differences between the records. 
Two of the widely used area catalogues of modern times, as produced by \citet{2002SoPh..211..357H} and \citet{2009JGRA..114.7104B}, utilize a combination of area observations from RGO and SOON (Solar Optical Observing Network). However, SOON data has several critical limitations. Sunspot area values in this catalogue, are significantly (by almost 50\%) underestimated as compared to RGO \citep{1997SoPh..173..427F,2002SoPh..211..357H,2009JGRA..114.7104B}. To a large extent, this is related to the fact that these data missed spots smaller than 10~$\mu$Hem and as the number of small spots varies with solar activity, a single calibration factor might introduce artefacts in the derived catalogues \citep[see][]{2014SoPh..289.1517F}. Furthermore, SOON has no direct overlap with RGO. Hence, the cross-calibration has to be done indirectly, e.g. using Russian data\footnote{from Russian books "Solnechnye Dannye"} as done by \citet{2009JGRA..114.7104B}, which amplifies the uncertainties further. Debrecen data whose area measurements are found to be similar to those from RGO \citep{2013MNRAS.434.1713B}, have a very short (3 years) overlap with RGO.

 Over the past few years, more sunspot data became publicly available in digital form. A significant development is that all data from the Pulkovo observatory (St. Petersburg) and its Mountain station in Kislovodsk have been digitised and made public \citep{Nagovitsyn1997}. These data are unique in the sense that ({\em i}) they cover a long period (1932--2018) allowing for a significant direct overlap with RGO, ({\em ii}) the smallest areas recorded in these catalogues are the same as in RGO, i.e. 1 millionth of a solar hemisphere ($\mu$Hem), and ({\em iii}) earlier studies \citep{1968SvA....11..976G,2009JGRA..114.7104B,2013MNRAS.434.1713B,2015ApJ...800...48M} showed that their statistical properties were very similar to those of the RGO data.
Also, daily sunspot observations from Kodaikanal solar observatory in India, have recently been digitised and cataloged \citep{2017A&A...601A.106M}. Similarly to Pulkovo and Kislovodsk, they cover a long period (1921--2011) and have a significant overlap with RGO.

 In this work, we update and extend the calibrated sunspot area series of \citet{2009JGRA..114.7104B} (hereafter BA09) by employing the additional and updated data sets. We describe the data we use in Sect.~\ref{section_data} and our methods in Sect.~\ref{section_method}. In Sect.~\ref{section_method} we present and discuss our composite records of sunspot areas i.e. daily corrected areas in Sect.~\ref{section_corrected}, daily projected areas in Sect.~\ref{section_projected} and individual group areas in Sect.~\ref{section_indi_area}.
In Sect.~\ref{section_psi}, we present the calculated daily PSI values (constructed using our area composite) which are an important input to empirical irradiance models. Our conclusions are summarized in Sect.~\ref{section_summary}. 

%-----------------------------------------
\begin{figure*}[!htbp]
\centering
\includegraphics[trim=0cm 0cm 0cm 2cm,clip,width=0.98\textwidth]{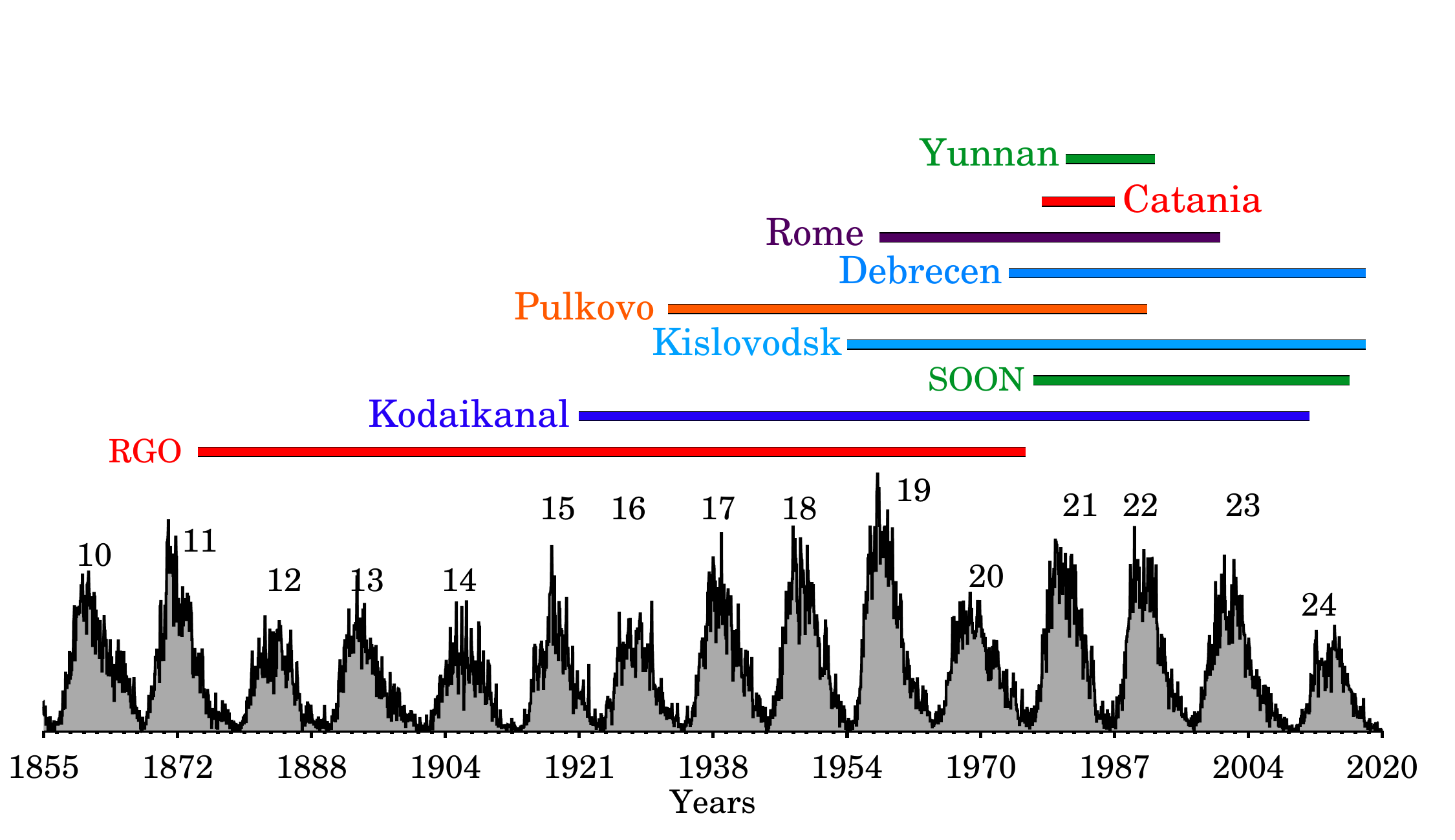}
\caption{Sunspot area datasets used in this work. Shaded curve in grey highlights the sunspot group number record for the reference period. See Table~\ref{event_details} for the abbreviations.} 
\label{context_image}
\end{figure*}
%-----------------------------------

%----------------------------------
\begin{table*}[!htb]
\begin{center}
\centering
\caption{  Details of each dataset used in this work}  
\vspace{0.2cm}
\label{event_details}
\setlength\tabcolsep{2.5pt}
\begin{tabular}{@{}llccccc@{}} 

  \hline

   Observatory & Abbr.  & Observing & Coverage  & Minimum reported &  available area type & Label \\
               &    &  period&         \%        &  Area  ($\mu$Hem)    &  (Projected / Corrected)    &   \\
     \hline
       RGO & RGO  & 1874-1976 & 99 & 1  & both & Primary\\
       Kislovodsk & Kisl  & 1952-2019 & 85 & 1 & both & Primary \\
       Pulkovo & Pul  & 1932-1991 & 86 & 1  & corrected & Primary\\
       Debrecen & Deb & 1974-2018$\dagger$ & 99 & 1 & both & Primary\\
       SOON & SOON  & 1981-2016 & 89 & 10 & both & Secondary\\
       Kodaikanal & Kodai & 1921-2011 & 58 & 2 & corrected &  Secondary\\
       Rome & Rome  & 1958-2000 & 50 & 2 & both & Secondary\\
       Catania & Cata  & 1978-1987 & 80 & 3 & both  & Secondary\\
       Yunnan & Yun  & 1981-1992 & 82 & 2 & both & Secondary\\

  \hline
\end{tabular}
\end{center}
\footnotesize{$\dagger$ 2016 onwards, Debrecen uses calibrated HMI observations to fill gaps.

Note: Projected areas refer to the observed values whereas corrected areas are adjusted for foreshortening effect.}
\end{table*}
%% -----------------------------

%====================DATA=====================
\section{Data} \label{section_data}

In this study we use sunspot area data from a total of nine observatories. Figure~\ref{context_image} shows the timeline of all these data sets, while Table~1 also lists the periods covered by each of them, the fractional temporal coverage and the minimum reported sunspot area.

The longest record comes from the Royal Greenwich Observatory (RGO) which started observing the Sun in 1874 and continued until 1976 \citep{2013SoPh..288..117W}. The observations have been carried out at several observatories at different locations (Royal Greenwich Observatory, England; Cape of Good Hope, South Africa; the Dehra Dun Observatory, India; the Kodaikanal Observatory, India; the Royal Alfred Observatory, Mauritius; along with contributions from the Harvard College Observatory; Melbourne Observatory; Mount Wilson Observatory and the US Naval Observatory), and then processed and combined into the final record at RGO. This allowed for an uninterrupted and consistent daily coverage over a period 100 years. This catalogue\footnote{\url{https://solarscience.msfc.nasa.gov/greenwch.shtml}} provides daily individual group areas as well as their heliographic positions. 

The next two datasets listed in Table~\ref{event_details}, are from Kislovodsk (1952--2018)\footnote{\url{http://158.250.29.123:8000/web/Soln_Dann/}} \citep{2007SoSyR..41...81N} and Pulkovo (1932--1991)\footnote{\url{http://www.gaoran.ru/database/csa/groups_e.html}} observatories \citep{1955Obs....75...28M}.
Pulkovo Observatory, originally established at 1839 with the aim of cataloging the positions of stars, started accumulating solar images (photosphere and chromosphere) in 1932. 
As in the case of RGO, observations were carried out at a number of various locations in the Soviet Union and then collected and processed at Pulkovo allowing for a consistency of the final series.
During the second world war, Pulkovo observatory was severely damaged, regular observations were not possible, and the original photographic plates from the pre-war period were destroyed. In 1945, the observatory received support from the government for restoration and continuation of the observational programme. Furthermore, the construction of a new branch, the Kislovodsk mountain station was initiated in 1948. Afterwards, both of these observatories, independently recorded daily sunspot data and their catalogues provide individual group area and positions.
 It is worth mentioning here that prior to May, 2011, the positional information (latitude and longitude) of each group is provided only for the day of its first appearance, while afterwards positions are available on each day throughout the entire lifetime of a group.

 The Debrecen observatory\footnote{\url{http://fenyi.solarobs.csfk.mta.hu/en/databases/DPD/}} is the official continuation of RGO programme since 1977 \citep{2011IAUS..273..403G,2016SoPh..291.3081B}. Most of the observations are taken at Debrecen observatory and its Gyula Observing 
Station. However, to fill gaps in this catalogue, observations from several contributing observatories (Abastumani Astrophysical Observatory, Georgia; Ebro Observatory, Spain; Helwan Observatory, Egypt; Kanzelh{\"o}ehe Solar Observatory, Austria; Kiev University Observatory, Ukraine; Kislovodsk Observing Station of Pulkovo Observatory, Russia; Kodaikanal Observatory, India and Tashkent Observatory, Uzbekistan) are also used. Recently (2016 onwards), the observatory started using calibrated SDO/HMI observations to fill the missing days in their catalogue.
In order to maintain consistency and also to avoid propagation of potential uncertainties due to this additional scaling, we only used the Debrecen data between 1974 and 2015 during our cross-calibration process.
We do, however, use the post-2016 Debrecen data to fill the remaining gaps (247 days have been filled with this data) in our final area composite after 2016 .

The newly digitized data from Kodaikanal solar observatory\footnote{\url{https://kso.iiap.res.in/new}} in India is next on our list. This set of newly digitized high resolution white-light solar images spans more than a century (1904--2011). However, due to issues with the observing plates, the current sunspot catalogue  lists data only from 1921 to 2011 \citep{2017A&A...601A.106M}. This catalogue provides individual spot areas and positions, however they are yet to be classified into sunspot groups.

We also use sunspot observations from SOON (Solar Optical Observing Network) \citep{2018SoPh..293..138G}. This is a network of solar observatories operated by the US Air Force (USAF), which allows a continuous, 24-h monitoring of the Sun. 
Finally, the last three data sets are those from: (i) Rome Astronomical Observatory\footnote{\url{ftp://ftp.ngdc.noaa.gov/STP/SOLAR_DATA/SUNSPOT_REGIONS/Rome/}} \citep{1967SoPh....2..375C}; (ii) Yunnan Observatory\footnote{\url{ftp://ftp.ngdc.noaa.gov/STP/SOLAR_DATA/SUNSPOT_REGIONS/Yunnan/}} in China \citep{1988sscd.conf..292W}; and (iii) Catania Astrophysical Observatory in Italy\footnote{\url{ftp://ftp.ngdc.noaa.gov/STP/SOLAR_DATA/SUNSPOT_REGIONS/Catania/}} \citep{1986soma.book.....D,2011CoSka..41...85Z}. All these three catalogues provide group areas and positions.
A summary of all the data used in this work is given in Table~\ref{event_details}.

Among all the sources (Table~\ref{event_details}), RGO has the longest observing period (about 100~years), the highest data coverage (99\%) as well as the smallest (together with Kislovodsk, Pulkovo and Debrecen) reported spot area (1$\mu$Hem). This makes RGO the most suitable as the reference series against which we calibrate all other records, as was also done by other studies in the past \citep{2002SoPh..211..357H,2009JGRA..114.7104B,2013MNRAS.434.1713B,2015ApJ...800...48M}. The remaining data sets have different quality as well as data coverage. The longest among the remaining records is from Kodaikanal (90 years), followed by Kislovodsk (68 years), Pulkovo (60 years), Debrecen (42 years), Rome (43 years) and SOON (36 years). Catania and Yunnan are relatively shorter records of roughly 10 years each.
Beside RGO, spots as small as 1~$\mu$Hem have also been reported by Kislovodsk, Pulkovo and Debrecen, whereas Kodaikanal, Rome, Yunnan and Catania recorded larger spots (see Table~\ref{event_details}). Finally, as mentioned above, SOON has a significantly higher lower area threshold of 10~$\mu$Hem. Earlier studies \citep{1968SvA....11..976G,2009JGRA..114.7104B,2013MNRAS.434.1713B,2015ApJ...800...48M} have shown that area measurements from Kislovodsk, Pulkovo and Debrecen are of mutually similar scale (having calibration coefficients close to 1). Therefore, we divide the listed sources into two categories, primary and secondary.
The purpose behind such labelling is to prioritize the ``primary" sources when creating a composite series later, while the secondary datasets should then be used to fill the gaps which could not be covered by the primary ones.
This classification is based on the following criteria. A dataset which is sufficiently long in time (3 solar cycles or more) and has comparatively few data gaps (i.e data coverage of 80\% or more), qualifies as a primary source. From Table~\ref{event_details} we notice that five sources: RGO, Pulkovo, Kislovodsk, Debrecen and SOON appear to satisfy these conditions. However, the minimum sunspot area reported in the SOON database, is significantly higher (10 $\mu$Hem) than the other four observatories (1 $\mu$Hem) and this can potentially affect the calibration process (see \citealp{2014SoPh..289.1517F}). Hence, SOON, along with the remaining four other observatories (Kodaikanal, Yunnan, Catania and Rome), get labelled as secondary sources.

%=======================METHOD===================
\section{Method} \label{section_method}

The method we adopted to cross-calibrate the individual records is similar to that described by BA09.
%---------------------------
\begin{figure*}[!htb]
\centering
\includegraphics[width=0.70\textwidth]{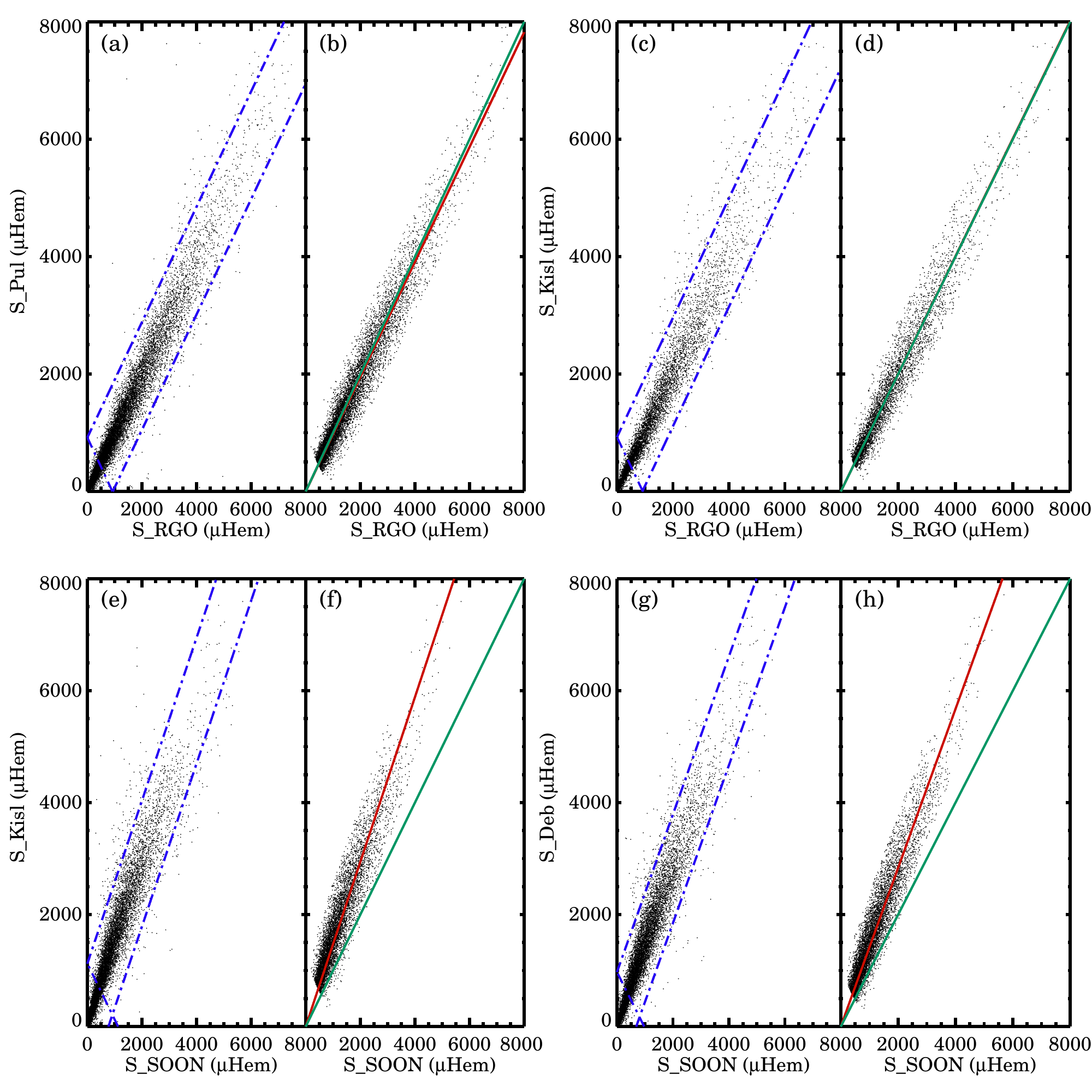}
\caption{Different steps of the calibration process for the following pairs of observatories: RGO-Pulkovo (panels-a,b); RGO-Kislovodsk (panels-c,d); SOON-Kislovodsk (panels-e,f); SOON-Debrecen (panels-c,d).  Blue dotted lines (in panels a, c, e, g) highlight the 3$\sigma$ boundaries which are used to remove the outliers and bias. Final `cleaned' scatter diagrams are plotted (in panels b, d, f, h) along  with the best linear fit (red lines) and the 45$^o$ slopes (green lines). See text for details.
}
\label{method_exp}
\end{figure*}
%--------------------------
 First, we identify the common observing days between any two observatories and perform the subsequent analyses over these overlapping periods only. We then plot a scatter diagram of daily area values between every given pair. Representative examples are shown in Fig.~\ref{method_exp}. We then fit a straight line forced to pass through the origin to the data: 
 \begin{equation}
\mathrm{X=b*Y},
\label{eqn1}
\end{equation}
 as shown by the solid red lines in Fig~\ref{method_exp}a, c, e, g.
 
 Points outside the 3$\sigma_{th}$ threshold, defined as
 \begin{equation}
 \sigma_{th}~=~\sqrt{\frac{1}{N-1}\sum_{i=1}^{N}\left( A_{i}^{Obs1}-b\cdot A_{i}^{Obs2}\right)^2 },
 \end{equation}
 from the previously obtained regression line are considered to be "outliers" and removed
 (highlighted with the blue dotted lines in Fig.~\ref{method_exp}a, c, e, g).
 We also reject points close to the origin (below the line joining the points [0, $3\sigma_{th}$] and [$3\sigma_{th}$, 0]) as they may introduce a bias into the calculated slope. At this stage, we are only left with the data which satisfy all the above criteria as shown in Fig.~\ref{method_exp}b, d, f, h. The linear regression (Eq.~\ref{eqn1}) is then applied once again to obtain a slope $\mathrm{b_{xy}}$. Since the choice of dependent and independent variable is completely arbitrary and may have an impact on the derived slope (commonly referred as the `attenuation bias'; \citealp{10.2307/1412159}), we repeat the above procedure by swapping the variables and obtain a new slope $\mathrm{b_{yx}}$. The final calibration factor ($\mathrm{``b"}$), following the `bisector line' method \citep{1990ApJ...364..104I}, is computed as
 \begin{equation}
 \mathrm{b}=(\mathrm{b_{xy}+1/b_{yx}})/2
 \end{equation}

The error associated with the final calibration factor ($\mathrm{``b"}$) has contributions from many different sources: (i) $\sigma_{slope}$: the fitting errors associated with the individual slopes $\mathrm{b_{xy}}$ and $\mathrm{b_{yx}}$; (ii)  $\sigma_{diff}$: the difference between the slopes $\mathrm{b_{xy}}$ and 1/$\mathrm{b_{yx}}$; and (iii)  $\sigma_{cycle}$: effects of the time dependent changes in the data onto the final $\mathrm{``b"}$ factor. For a chosen pair, we compute $\rm{``b"}$ for different solar cycles separately and the standard deviation of these individually measured $\rm{``b"}$ values is taken as $\sigma_{cycle}$. 
Thus, the final uncertainty, rather conservatively, is calculated as $\sigma=\sqrt{{\sigma_{slope}}^2+{\sigma_{diff}}^2+{\sigma_{cycle}}^2}$.

%=================RESULTS=============================
\section{Results and discussion}\label{section_results}

%++++++++++++++++CORRECTED AREAS++++++++++++++++++++++
\subsection{Corrected areas}\label{section_corrected}
\subsubsection{Comparison of individual records}\label{section_indi_comp}

We apply the method described in Section~\ref{section_method} on all pairs of observatories that have an overlap with each other. The derived parameters, obtained using the corrected areas, are tabulated in Table~\ref{cal_table}. The table lists individual calibration factors $\rm{b}_{xy}$ and $\rm{b}_{yx}$ (columns four and five) as well as the final calibration factor $\rm{``b"}$ (last column). We multiplied the areas recorded by observatories listed under `Obs2' with $\rm{``b"}$ to match the values from those listed under `Obs1'. With this definition, values of $\rm{``b"}$ close to 1 imply that the original area measurements obtained at the two observatories, on average, are similar to each other.

Let us first discuss those cases where an observatory has a direct overlap with RGO (this is the case for Kislovodsk, Pulkovo, Rome and Kodaikanal). The $\rm{``b"}$ factor between RGO and Pulkovo, $\rm{b_{RGO-Pul}}$ is derived to be 1.014$\pm$0.069, in agreement with BA09's result of 1.019. Similarly, $\rm{b_{RGO-Kisl}}$ comes out to be 0.984$\pm$0.094, which also agrees with the value of 0.979 reported by \citet{2013MNRAS.434.1713B}. Thus, area measurements from Pulkovo and Kislovodsk are similar to those from RGO. Furthermore, \citet{2015ApJ...800...48M} found that the individual sunspot group size distributions are also similar to each other. This is important for building a composite area series. The situation is different when we compare RGO to Kodaikanal. We find $\rm{b_{RGO-Kodai}}$ to be 1.166$\pm$0.132, which indicates that the spot areas in the current Kodaikanal catalogue are lower ($\approx$17\%) than the RGO values. This can also be seen (at least qualitatively) in \citet{2017A&A...601A.106M}. Between RGO and Rome, $\rm{b_{RGO-Rome}}$ equals to 1.091$\pm$0.036 which is similar to the value obtained by BA09.

%---------------------------
 \begin{table}
\begin{center}
\centering
\caption{Calibration factors derived for different observatories.}  
\vspace{0.3cm}
\label{cal_table}
\setlength\tabcolsep{2.5pt}
\begin{tabular}{llccccc}

  \hline

   Obs1 [y] & Obs2 [x] & Overlap & $\rm{b_{xy}}$ & $\rm{b_{yx}}$ & b & CC\\
     
     \hline
       RGO & Kisl & 1954--1976 & 0.969 & 1.001 & 0.984$\pm$0.094 & 0.971 \\
       RGO & Pul & 1932--1976 & 0.998 & 0.977 & 1.014$\pm$0.069 & 0.972\\
       RGO & Kodai & 1921--1976 & 1.099 & 0.811 & 1.166$\pm$0.132 & 0.857\\
       RGO & Rome & 1958--1976 & 1.073 & 0.901 & 1.091$\pm$0.036 & 0.968\\
       Deb & Kisl & 1974--2016 & 0.944 & 1.028 & 0.958$\pm$0.028 & 0.962\\
       Deb & Pul & 1974--1991 & 0.921 & 1.058 & 0.932$\pm$0.026 & 0.966\\
       Kisl & Pul & 1954--1991 & 1.003 & 0.984 & 1.009$\pm$0.012 & 0.984\\
       Kisl & Yun & 1981--1992 & 1.303 & 0.731 & 1.335$\pm$0.063 & 0.930\\
       Pul & Rome & 1958--1991 & 1.109 & 0.854 & 1.140$\pm$0.061 & 0.905\\
       Kisl & SOON & 1977--2016 & 1.471 & 0.650 & 1.504$\pm$0.066 & 0.942\\
       Kisl & Kodai & 1954--2011 & 1.048 & 0.853 & 1.109$\pm$0.123 & 0.823\\
       Rome & Catania & 1978--1987 & 0.961 & 0.971 & 0.995$\pm$0.069 & 0.881\\

       RGO & Deb & Via Pul & -- & -- & 1.061$\pm$0.091 & --\\
       RGO & SOON & Via Kisl & -- & -- & 1.480$\pm$0.102 & -- \\
  \hline

\end{tabular}
\end{center}
\end{table}
%% -------------------------

 Next, we look at those observatories which have an insignificantly short ( less than several years) overlap with RGO. Their inter-calibrations are accomplished `indirectly' i.e. by using another source which overlaps with these observatories with RGO. Two of the longest series in this list are Debrecen and SOON. Using the overlap between Debrecen and Pulkovo, we calculate $\mathrm{b_{RGO-Deb}}$ to be 1.061$\pm$0.091, which is consistent with the factor of 1.08 reported by \citet{2013MNRAS.434.1713B}, even though it was obtained from a shorter period of Debrecen data.
 
 For SOON, $\mathrm{b_{RGO-SOON}}$ is estimated via Kislovodsk and is found to be 1.48$\pm$0.102. Once again, it matches the values previously reported in the literature \citep{2002SoPh..211..357H,2009JGRA..114.7104B}. It is important to note that among all our data sources, SOON has the maximum area departure from RGO (almost 50\%). As discussed in the introduction, there can be a number of reasons for this significant underestimation. According to \citet{2014SoPh..289.1517F}, it is those {\it `too small to draw'} spots ($<$10$\mu$Hem) in the SOON catalogue which are mostly responsible for this deficit. However, \citet{2017MNRAS.465.1259G} argued that the omission of small spots can only account for $\approx$3.4\% of the area deficit and the measurement procedure may be responsible for the rest. By reanalyzing a portion of SOON data, \citet{2018SoPh..293..138G} concluded that the rounding errors associated with the limb-correction overlay, used on the SOON drawings, can actually lead to an underestimation of spot areas as much as 8.5\%.

%--------------------------
\begin{figure}[!htb]
\centering
\includegraphics[width=0.45\textwidth]{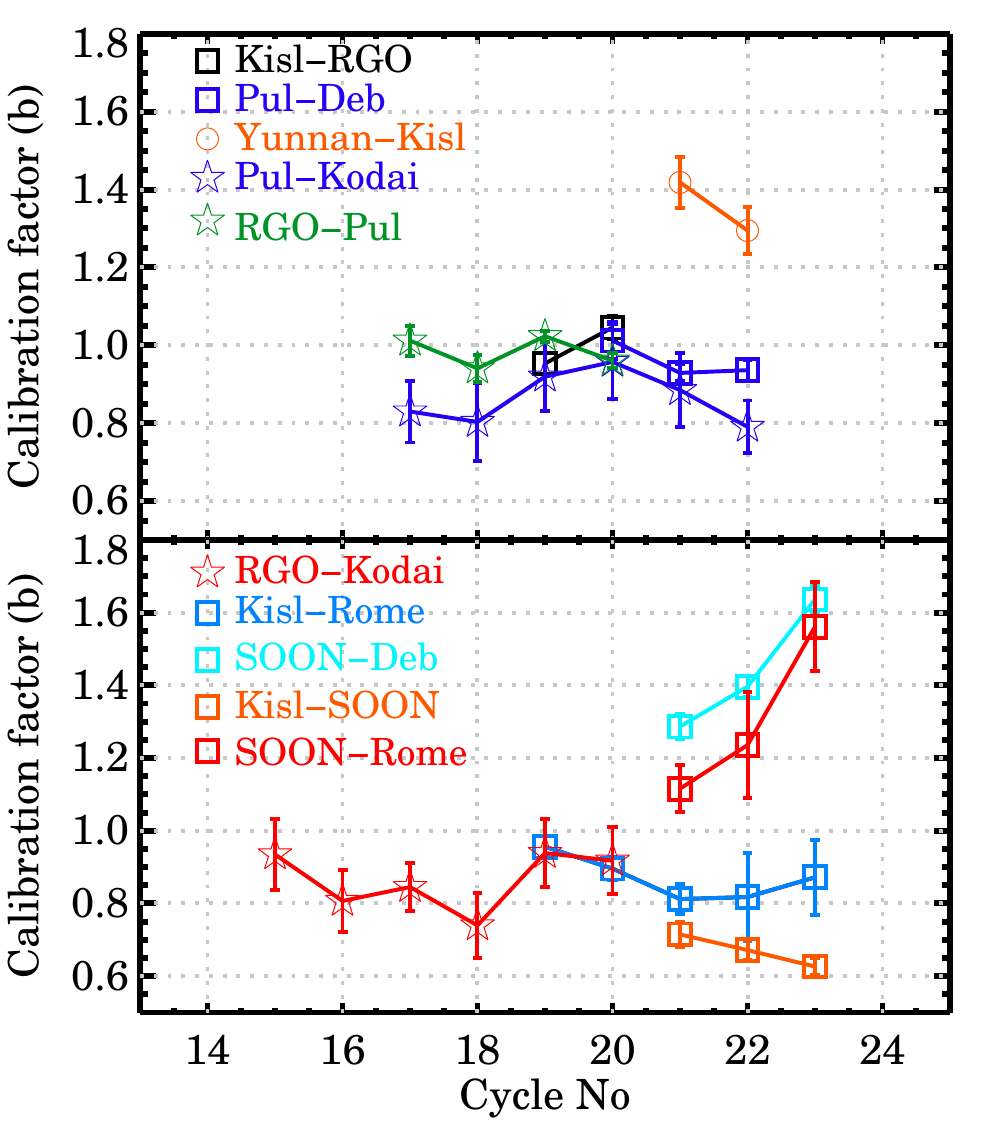}
\caption{The value of the calibration factor $\mathrm{``b"}$ between various pairs of observatories (see the legends in the panels) computed for each solar cycle separately.} 
\label{cycle_var}
\end{figure}
%-----------------------
One of the main issues in calibrating areas between two observatories, is to address the temporal evolution within a dataset. These fluctuations can arise due to changes in quality of instruments or capturing devices and measuring techniques, as well as from aging due to preservation of sunspot drawings and photographic plates over a longer time. Now, any such changes in one or both series will show up as time evolution in the derived calibration factor.
 To see the extent of such an effect, we plot the values of $\rm{``b"}$, computed for each cycle separately, as a function of time in Fig.~\ref{cycle_var}.

Variations over shorter timescales (monthly or yearly) are not considered here as they are significantly affected by uncertainties coming from insufficient statistics.
Different lines in  Fig.~\ref{cycle_var} with various colors and symbols represent the evolution of $\rm{``b"}$ for different pairs of observatories (see legend in the Figure). 

Figure~\ref{cycle_var} demonstrates that $\rm{``b"}$ does vary with time for all tested combinations of observatories.
However, for the cases, when both data sets in question are our "primary" choice (see Sect. 2 and Table 1), the variations are within the error-bars. 
In some other cases, the calibration factor shows significantly larger variations, e.g between SOON and Debrecen or Kodaikanal and RGO.
However, all cases with large fluctuations (e.g., where Kodaikanal or SOON data enter) are found for the secondary sources which were merely meant to be used to fill occasional data gaps. 
This result (1) supports our choice of the primary sources and (2) justifies the
use of a single $\rm{``b"}$ value for each pair of observatories (as listed in Table~\ref{cal_table}) for building the composite record. 

%+++++++++++++++++COMPOSITE++++++++++++++++++++++++
\subsubsection{Composite}\label{section_corrected_composite}
\begin{figure*}[!htb]
\centering

\includegraphics[trim = 0.4cm 0mm 0.7cm 0.5cm, clip,width=0.85\textwidth]{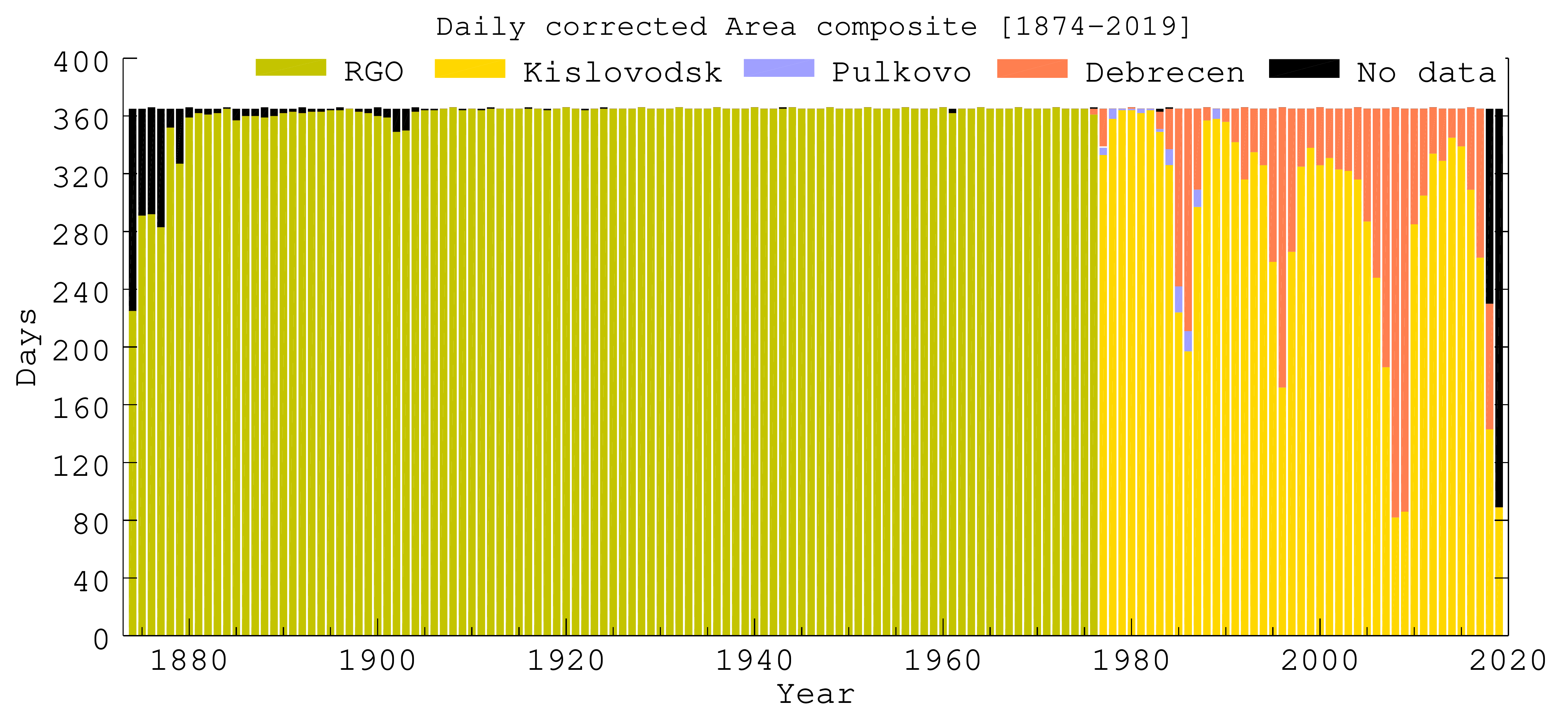}
\hbox{\hspace{1.2cm} \includegraphics[trim = 0.6cm 0mm 9cm 2cm, clip,width=0.75\textwidth]{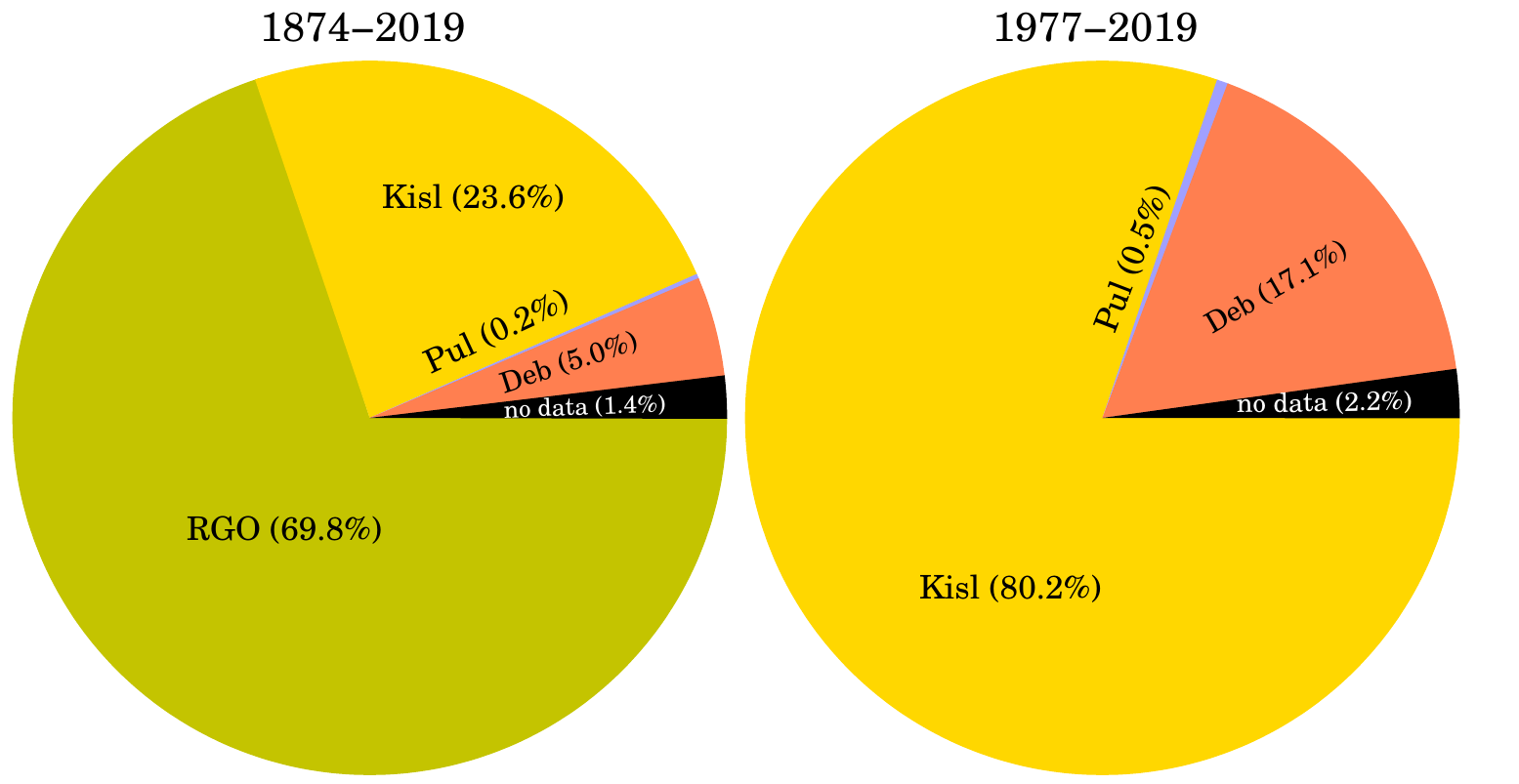}}

\caption{
Top panel: An overview of the structure and the coverage of the final composite of the corrected sunspot areas. Different colours (see legend at the top) show data from different observatories. Y-axis is the number of days per year, for which data are available. Bottom panel: Pie-charts highlighting the percentage of contributions of observatories to the complete calibrated series (1874-2019: left chart) and only to the post-RGO period (1977-2019: right chart).
} 
\label{timeline_graphic }
\end{figure*}

At this stage, we are ready to generate a calibrated and homogeneous sunspot area series  between 1874 and 2019.
We start by using the data from four primary observatories from our list (Table 1), i.e RGO, Pulkovo, Kislovodsk and Debrecen. 
RGO, being the absolute reference (for the reasons discussed in Sec.~\ref{section_data}), is used as it is.
Next, both Kislovodsk and Pulkovo have a direct and sufficiently long overlap with RGO (which Debrecen does not have).
Their $\rm{``b"}$ values are also similar (see Table~\ref{cal_table}). 
However observations from Kislovodsk are considered to be better suited for the extension of RGO because of the stable background history of this catalogue \citep{2007SoSyR..41...81N}.
The other advantage of the Kislovodsk record over Pulkovo is that it offers an additional 28 years of added observations beyond 1991.
Hence, we use areas from Kislovodsk as the main record to extend our catalogue after the RGO period. 
The leftover missing days are first filled with areas from Pulkovo and then from Debrecen (see Fig.~\ref{timeline_graphic } and Table~\ref{correted_summary_table} for
a summary of the observations constituting the final composite of daily corrected areas). 
 
%-------------------------------
 \begin{table}
 \begin{center}
 \centering
 \caption{Summary of the observations used for the final daily corrected area composite. Last column of the table, lists the overall fraction of the data from a given archive used for the composite. For example, all the available RGO data (100\%) have been used in the final catalogue, whereas only 16\% of the available Debrecen data are used.}
 \vspace{0.3cm}
 \label{correted_summary_table}
 \begin{tabular}{cccc} \hline
     Order   & Obs.  & Data & \% of which, used  \\
      &   name       & coverage    &  in this catalogue   \\
     \hline
      1     & RGO  & 1876-1976 & 100\% \\
      2     & Kislovodsk & 1952-2019 & 62\% \\
      3     & Pulkovo & 1932-1991  &   0.4\% \\
      4     & Debrecen & 1974-2018  & 16\% \\
     \hline

  \end{tabular}
  \end{center}
\end{table}
%% ---------------------------

\begin{figure*}[!htb]
\centering
\includegraphics[width=0.80\textwidth]{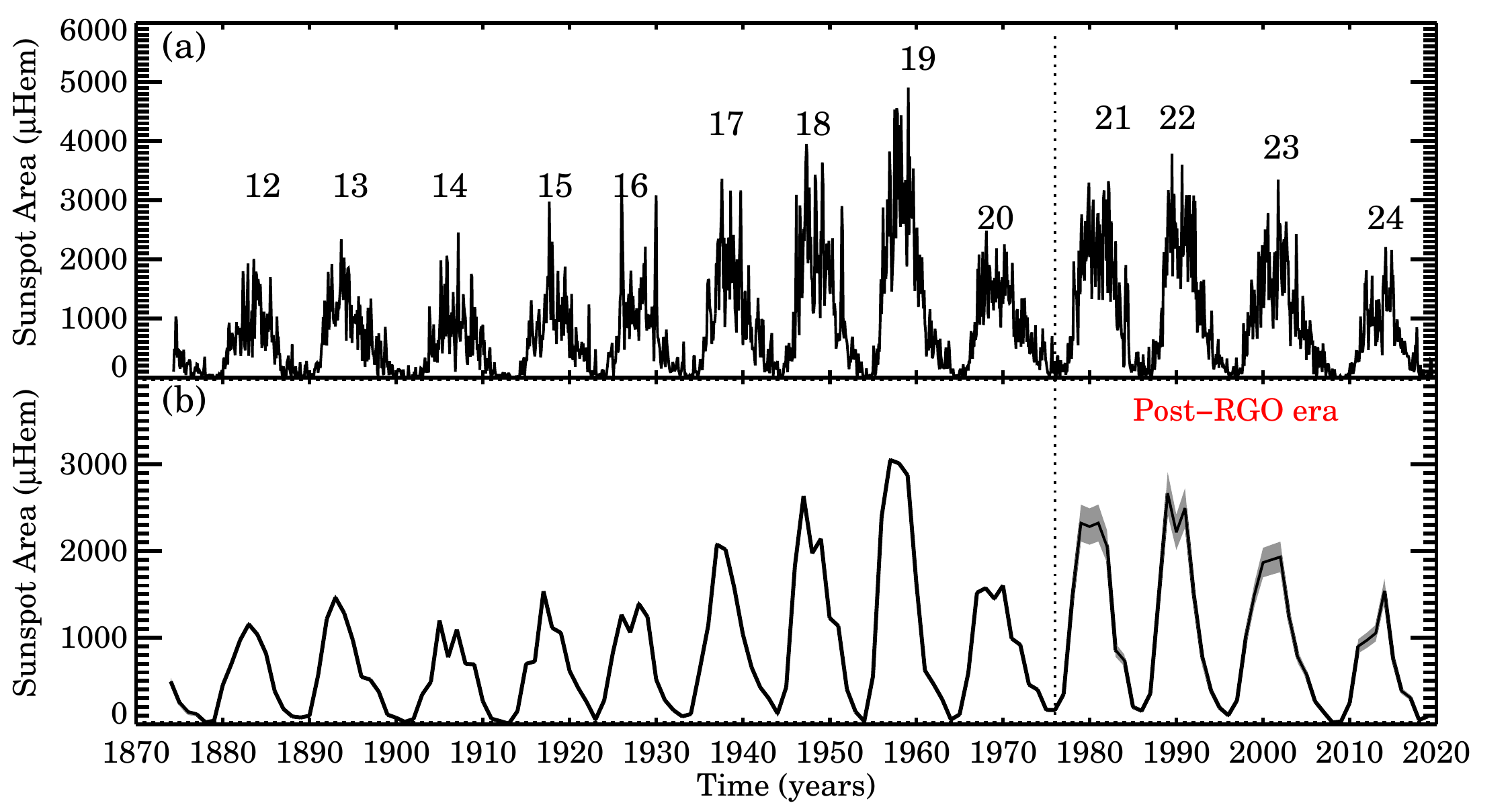}
\caption{ Monthly (panel-a) and yearly (panel-b) averaged calibrated sunspot area series. Calculated error values (grey shaded regions) are only shown for the yearly series. The dotted vertical line marks the year 1976, when RGO stopped its observing campaign. } 
\label{cal_area}
\end{figure*}

Our final catalogue contains about 145 years of daily sunspot area values between 1874 and 2019. This catalogue is available online with this publication and at \url{http://www2.mps.mpg.de/projects/sun-climate/data.html}. The total number of missing days in this series is 776 (corresponding to 1.4\% of the total coverage). We could not fill these missing days with data from any of the remaining five observatories (Kodaikanal, Rome, Yunnan, SOON, Catania) because out of the 776 missing days, 443 days are between 1874--1922 where only RGO observations are available and 321 days are between July, 2018~-- Dec, 2019 where only observations from Kislovodsk are available (Figure~\ref{timeline_graphic }). We note here that the cataloging process at Kislovodsk and Debrecen for the last two years (2018 onward) is still in progress and we plan to fill these missing days as soon as the data become available.
While we have compared a total of nine archives, only four of them have actually entered the final composite. We nevertheless show the results obtained from inter-comparisons of these ``secondary'' datasets  and list their scaling coefficients in Table~\ref{cal_table} for completeness.
Panels a and b of Fig.~\ref{cal_area} show the calibrated monthly and yearly averaged time series of corrected areas.
To visualize the uncertainty, we overplot two area series generated with the two extreme limits of the errors in $\rm{``b"}$ i.e $\mathrm{b+\sigma}$ and $\mathrm{b-\sigma}$ (from Table~\ref{cal_table}), shown as the shaded regions in Fig.~\ref{cal_area}b.
As expected, the effect is prominent mostly during solar maxima when the total spot coverage is higher. This results in the corresponding uncertainty 
in the cycle amplitudes over the post-RGO period, which has to be kept in mind in relevant studies.

%---------------------------------
%------------------------------

%---------------------------------
%------------------------------

%+++++++++++++++++++++++COMPARISON++++++++++++++++++
\subsubsection{Comparison with BA09} \label{section_area_comp}

While we have generally followed the procedure by BA09, there are also some differences. (i) Firstly, 
instead of SOON we use Kislovodsk and Debrecen data for the post-RGO period. With Kislovodsk data we extend our series till 2019 (the BA09 series ended in 2009), while with Debrecen data, we improve the daily data coverage by filling most of the intermittent data gaps. 
(ii) Secondly, our all four observatories (Kislovodsk, Pulkovo, Debrecen) have calibration factors ($\rm{``b"}$) close to 1 whereas for the SOON data, used by BA09\footnote{BA09 data is downloaded from \url{https://agupubs.onlinelibrary.wiley.com/doi/full/10.1029/2009JA014299}}, the value of $\rm{``b"}$ is $\approx$1.5. Hence, the uncertainties are expected to be lower in our catalogue.  
(iii) Next, since RGO and SOON do not overlap directly, BA09 employed published Russian data for their cross-calibration.
We use data from Kislovodsk and Pulkovo to extend the RGO series and both of them have significant overlaps with RGO. It is worth mentioning that the `Russian data' used by BA09 started only in 1968 whereas the updated Pulkovo catalogue which we use here goes back to 1932. This significantly increases the overlap with RGO, which again helps to minimize the uncertainties.

%------------------------------------
\begin{figure*}[!htb]
\centering
\includegraphics[width=0.80\textwidth]{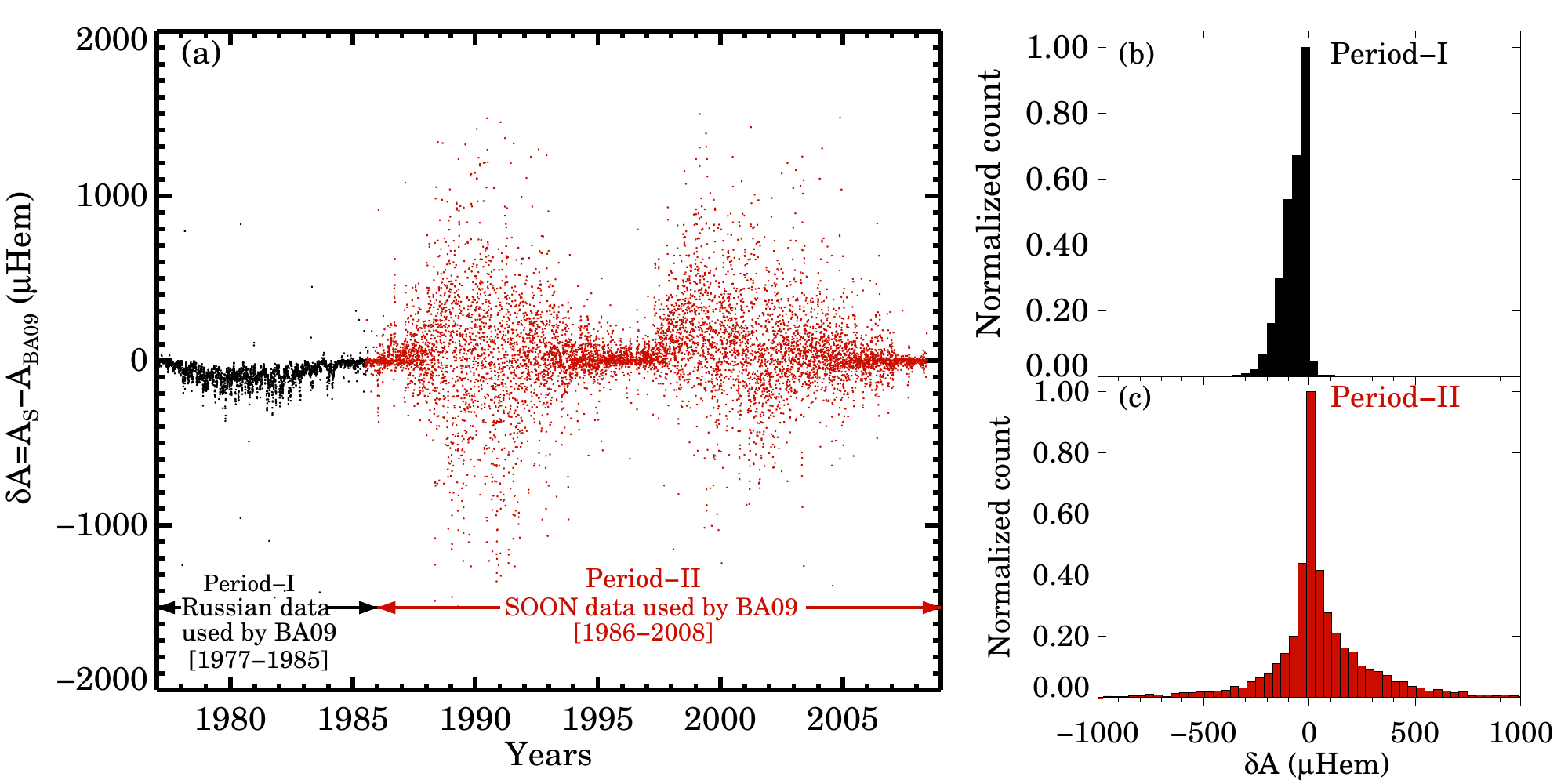}
\caption{Panel(a): Differences ($\delta \rm{A}$=$\rm{A_S}$-$\rm{A_{09}}$) between daily area values obtained in this study ($\rm{A_S}$) and those published by BA09 (A$_{09}$). Panel(b) and (c): Histograms of $\delta\rm{A}$ for the two periods (Period-I: 1977-1985; Period-II: 1986-2008).} 
\label{compare_plot}
\end{figure*}
%-----------------------------------
%------------------------------
 \begin{table}
\begin{center}
\centering
\caption{ Examples of area values when $\delta$A is $\ge$1000 $\mu$Hem}  
\vspace{0.3cm}
\label{compare_table}
\setlength\tabcolsep{2.5pt}
\begin{tabular}{lccccc}

  \hline

     yyyymmdd & This\_study  & BA09 & Deb & Kodai & Rome  \\
              & Kisl$\times$0.98  & SOON$\times$1.49 & $\times$1.06 & $\times$1.17 & $\times$1.09  \\

     \hline
    19881025 &   4035   &     2106  &      2420  &   2822   &     2211\\
    19880628 &   3532   &     4772  &      3641  &   5289   &     2166\\
    19890831 &   3673   &     4890  &      4736  &   3774   &     3538\\
    19901115 &   2974   &     4678  &      4703  &   4364   &     4389\\
    19901119 &   3010   &     4168  &      3699  &   3994   &     4597\\
    19990731 &   2832   &     1335  &      2908  &   2403   &     --\\
    19900927 &   1375   &     2591  &      1171  &   2011   &    1178\\
    19910130 &   6565   &     7594  &      6942  &   6840   &    6687\\
    19950418 &   624    &     2816  &      461   &   535    &    323\\
    19990604 &   2692   &     1468  &      2797  &   --     &    2338\\
  \hline

\end{tabular}
\end{center}
\footnotesize{Note: All areas are in the unit of $\mu$Hem.}
\end{table}
%% ------------------------------

Let us now compare the two compilations quantitatively. Since the RGO dataset is essentially the same in both studies, we focus only on the post-RGO era, i.e. between 1977 and 2008 (when the BA09 series ended). In this period, our catalogue utilises daily data ($\rm{A_S}$) from Kislovodsk whereas BA09 used observations ($\rm{A_{09}}$) from Russian books "Solnechnye Dannye" (Period-I; between 1977--1985) and SOON (Period-II; between 1986--2008). The daily difference between the two composites, $\delta \rm{A}$=$\rm{A_S}$-$\rm{A_{09}}$, is plotted in Fig.~\ref{compare_plot}a. We also separately plot the histograms of $\delta\rm{A}$ for the two periods in Figure~\ref{compare_plot}b and Figure~\ref{compare_plot}c. 
As seen from the figure, for Period-I (black dots), our area values are systematically lower (by $\sim$6\%) compared to areas in BA09. This was, in fact, already noticed by BA09, who reported that the Russian area measurements used in their study were systematically larger than RGO by $\sim$8\% between 1971 and 1976.
However, without being able to do a detailed analysis of the reasons for this change of the correction factor with time, they refrained from correcting it.
The Kislovodsk data that we use here do not show such an offset, and thus solve this issue with the compilation of BA09. For Period-II (red dots), BA09 used data from SOON whereas out catalogue uses Kislovodsk areas throughout. For this period, we do not see any systematic drift, but rather the differences are distributed symmetrically, with most of the values ($\sim$80\%) being below 200\,$\mu$Hem (Fig.~\ref{compare_plot}). The differences are clearly higher during higher-activity periods, when the number of spots is considerably larger.

 Now, these smaller differences ($\le$ $\mid 200\mid$ $\mu$Hem) are rather difficult to diagnose due to various uncertainties in area measurements as well as in the analysis procedure. But let us take a closer look at those days where the absolute difference is more than 500 $\mu$Hem (although such cases are rather rare, $<$8\%).
As mentioned already, in this period BA09 used data from SOON whereas we use data from Kislovodsk. To better identify the source of the discrepancies, we also compare measurements on the same days from three other observatories: Debrecen, Kodaikanal and Rome. A small sub-sample (showing extreme departures of $\delta\rm{A}$ $\ge$1000 $\mu$Hem) is presented in Table~\ref{compare_table}.

After comparing the area values across observatories on various days, we could identify essentially all possible scenarios. From $\rm{A_{S}}$ or $\rm{A_{09}}$ matching with no other records to both of them match only with some. This can be also be presented quantitatively by using a set of tolerance values (which account for the possible measurement errors in the original datasets). Comparing area values in this way, we find that, for $\sim$ 50\% to 70\% of cases, either $\rm{A_{S}}$ or $\rm{A_{09}}$ are the single outliers. In roughly 30\% to 50\% of the cases, at least one other observatory provided a value similar to either $\rm{A_{S}}$ or $\rm{A_{09}}$. These results show that there is no systematic bias towards one of the data sets, e.g., Kislovodsk, Debrecen, SOON, etc. A more sophisticated and robust technique, e.g. a `spot-to-spot’ calibration, is needed to address and correct for these problems. This is beyond the scope of this current work and needs a separate study.

%------------------------------
\begin{figure*}[!htb]
\centering
\includegraphics[width=0.85\textwidth]{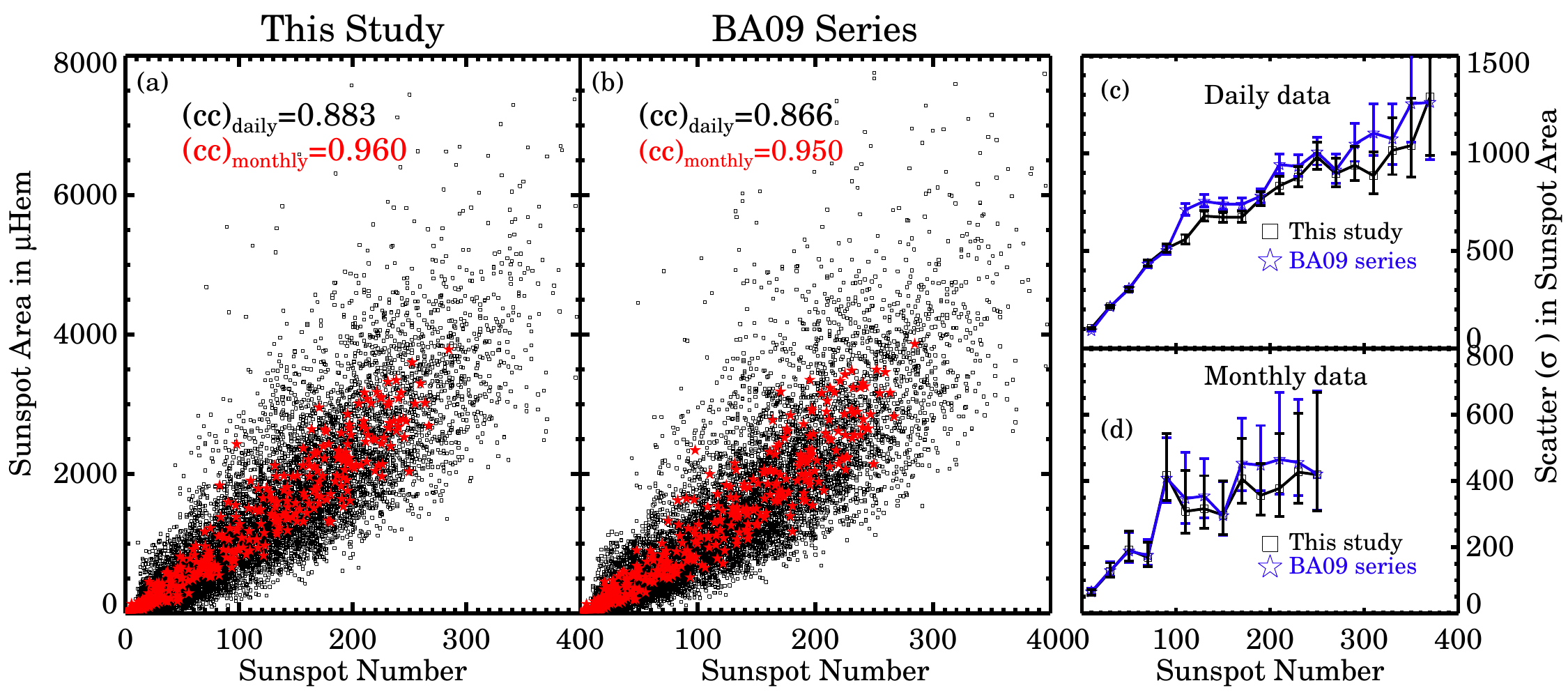}
\caption{a: Scatter diagram between daily (monthly) sunspot number and daily (monthly) sunspot areas from this work, in black (red). b: The same but for the areas from the BA09 series. The relevant correlation coefficients (cc) are printed in the respective panels. c and d: Binned values of scatter ($\sigma$) in sunspot areas (this work in black and BA09 series in blue) vs. sunspot number (binned values of 20) for the daily and monthly data, respectively. Error-bars represent the 90\% confidence intervals of $\sigma$.} 
\label{sunspot_number_area}
\end{figure*}
%------------------------------

We have also compared the two composites with the sunspot number series\footnote{Sunspot number series V2.0 from SIDC; \url{http://www.sidc.be/silso/datafiles}}. We only compare the post-RGO period (1977-2008). Panel~\ref{sunspot_number_area}a shows a scatter diagram between daily (monthly) sunspot number and daily (monthly) sunspot areas from this work, in black (red). The same but using the BA09 series is shown in panel~\ref{sunspot_number_area}b. The Pearson correlation coefficients ($R_c$) for daily records is $R_{c, this\_work}$=0.883 vs $R_{c, BA09}$=0.866 and for monthly data $R_{c, this\_work}$=0.960 vs $R_{c, BA09}$=0.950. Allowing for the non-linear relation between sunspot number and area, we also calculate the Spearman's rank correlation coefficients ($\rho$) for daily records, $\rho_{\_this\_work}$=0.934 vs $\rho_{\_BA09}$=0.934, and for monthly data, $\rho_{\_this\_work}$=0.971 vs $\rho_{\_BA09}$=0.960.
We also compute the scatter (as the standard deviation, $\sigma$) in the area values within the bins of 20 in the sunspot number as well as the 90\% confidence intervals of the $\sigma$. Panels c and d of Fig.~\ref{sunspot_number_area} show these results for the daily and monthly data, respectively. The scatter in the daily values (panel~c) in our series is lower than that in BA09 for a significant part of the sunspot number range. The scatter in the monthly data (panel~d) is comparable in the two series.

%++++++++++++++++PROJECTED AREAS+++++++++++++++++
\subsection{Projected areas} \label{section_projected}
 
%---------------------------
\begin{figure*}[!h]
\centering
\includegraphics[width=0.80\textwidth]{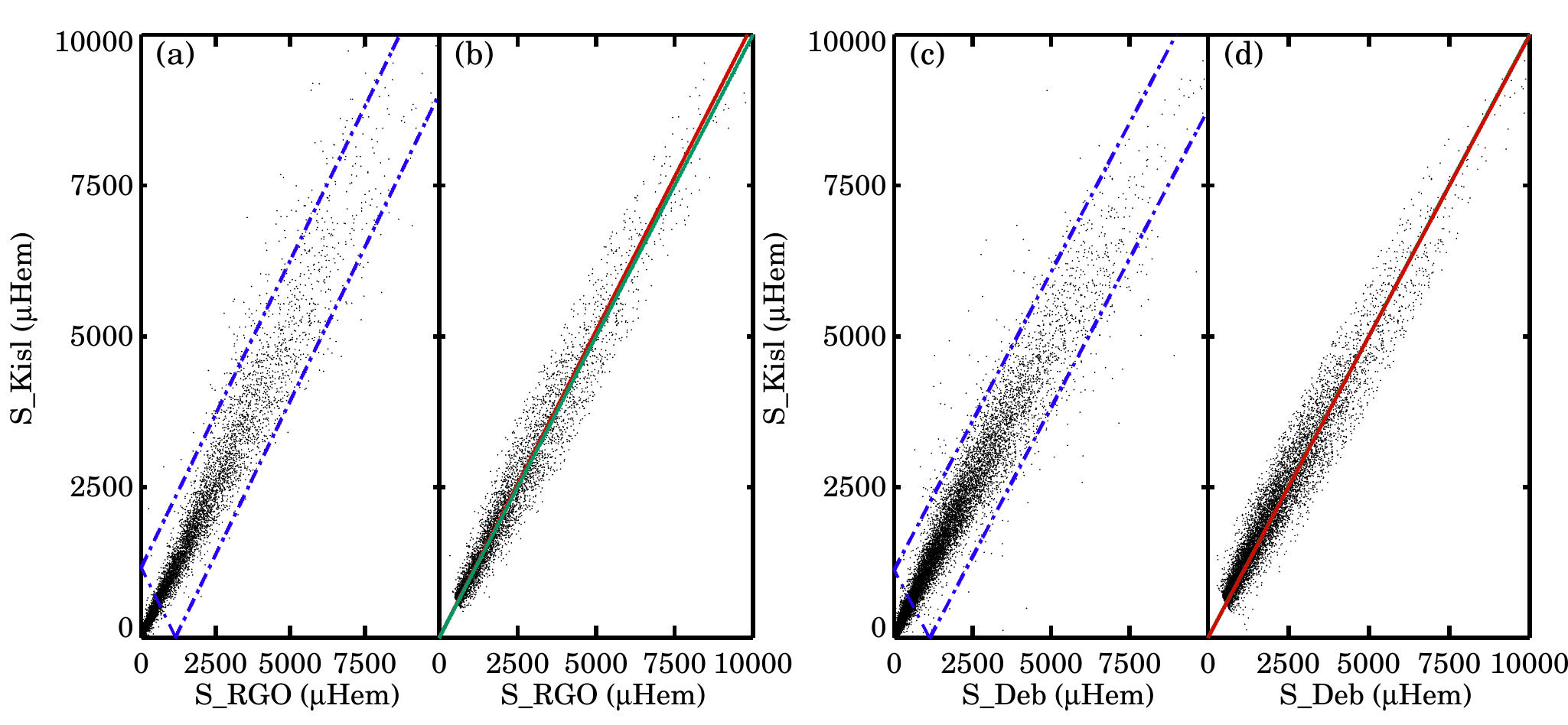}
\caption{Similar to Figure~\ref{method_exp}, but generated using projected area values. Blue dotted lines (in panels-a,c,) highlight the 3$\sigma$ boundaries which are used to remove the outliers and bias. Final `cleaned' scatter diagrams are plotted (in panels-b,d) along  with the best linear fit (red lines) and the 45$^o$ slopes (green lines). } 
\label{appendix_1}
\end{figure*}
%---------------------------

Studies such as irradiance reconstruction \citep{2010JGRA..11512112K,2017JGRA..122.3888Y}
use the projected area values as an input. Hence, in addition to the corrected area values, we also perform the cross-calibration with the projected areas. To achieve this, we use the same set of primary observatories as used in corrected area composite before, except Pulkovo.
Pulkovo does not provide the projected areas and only lists the corrected ones (Table~\ref{event_details}). This would not have been an issue had Pulkovo provided the time of observations which are required to transform the corrected areas into projected ones (as the longitudes are listed in Carrington coordinates). 
Hence, we decided to leave out data from Pulkovo and only use data from RGO, Kislovodsk and Debrecen to generate this catalogue.
The method of cross-calibrations is the same as described previously and the results are plotted in Figure~\ref{appendix_1}. Derived calibration factors are, $\rm{b_{RGO-Kisl}}$=1.02$\pm$0.025 and $\rm{b_{Deb-Kisl}}$=1.01$\pm$0.026. A summary of the final calibrated series of daily projected areas is given in Table~\ref{projected_summary_table}. 

%-------------------------------
 \begin{table}[!h]
 \begin{center}
 \centering
 \caption{Summary of the observations used for the final daily projected area composite. Last column of the table, lists the overall fraction of the data from a given archive used for the composite. For example, all the available RGO data (100\%) have been used in the final catalogue, whereas only 16\% of the available Debrecen data are used.}
 \vspace{0.3cm}
 \label{projected_summary_table}
 \begin{tabular}{cccc} \hline
     Order   & Obs.  & Data & \% of which, used  \\
      &   name       & coverage    &  in this catalogue   \\
     \hline
      1     & RGO  & 1876-1976 & 100\% \\
      2     & Kislovodsk & 1952-2019 & 62\% \\
      3     & Debrecen & 1974-2018  & 17\% \\
            \hline

  \end{tabular}
  \end{center}
\end{table}
%% ---------------------------

%================INDI AREAS====================
\subsection{\textbf{Individual group areas}}\label{section_indi_area}

 %-----------------------------------
   \begin{figure*}[!h]
    \centering
    \includegraphics[width=0.90\textwidth]{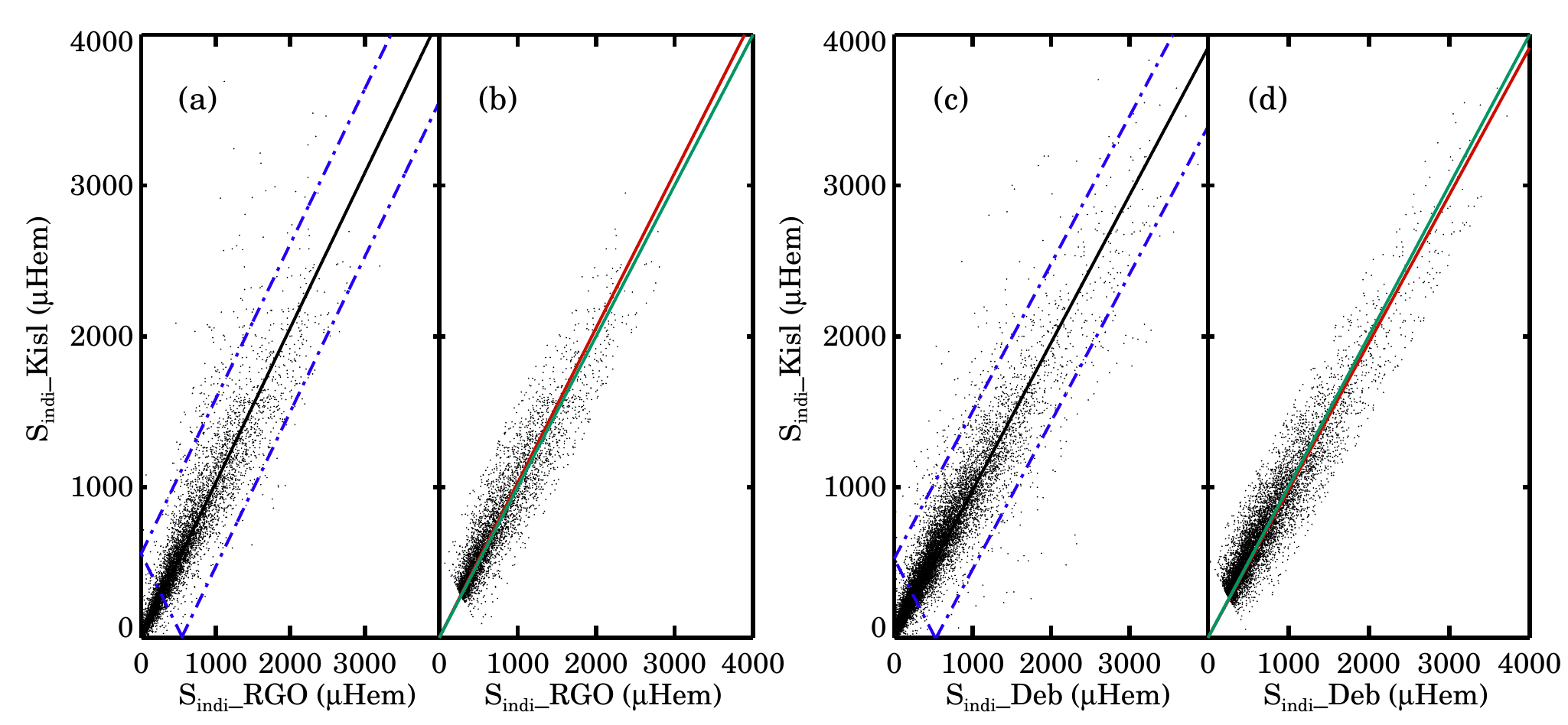}
     \caption{Similar to Figure~\ref{method_exp}, but generated using corrected individual group areas. Blue dotted lines (in panels-a,c,) highlight the 3$\sigma$ boundaries which are used to remove the outliers and bias. Final `cleaned' scatter diagrams are plotted (in panels-b,d) along  with the best linear fit (red lines) and the 45$^o$ slopes (green lines). } 
     \label{indi_area_comp} 
  \end{figure*}
%---------------------------------------

Some applications, such as the Surface Flux Transport (SFT) models often used to reconstruct the evolution of the surface magnetic fields and irradiance \citep[see, e.g.,][]{2011A&A...528A..83J,2014SSRv..186..491J}, it is important to have information on individual groups.
 Hence, in addition to the daily calibrated areas, we also provide the individual group areas. 
 A direct comparison of individual sunspot groups among multiple datasets is not a trivial task.
 This requires not only an identification of the same group across different datasets, but also accounting for the group evolution due to the difference in observing times between observatories. This itself is a subject of a separate study and is beyond the scope of this current paper. Nonetheless, since this information is needed, we perform a simple comparison here without such a detailed study and provide a preliminary record of individual group areas.

  For this purpose, we use individual group areas (corrected for foreshortening) from RGO, Kislovodsk and Debrecen and only choose the biggest individual group per day in each of these three observatories. The rest of the analysis is the same as presented in Sect.~\ref{section_method}.
   The results for  RGO vs. Kislovodsk and Debrecen vs. Kislovodsk are shown in Fig.~\ref{indi_area_comp}a,b and Fig.~\ref{indi_area_comp}c,d, respectively.
   The derived calibration factors are, $\rm{b_{RGO-Kisl}}$=1.031$\pm$0.056 and $\rm{b_{Deb-Kisl}}$=1.006$\pm$0.046.
   They are similar to the ones we previously obtained with the daily corrected areas (Table~\ref{cal_table}). Thus, this preliminary analysis suggests that the calibration factors listed in Table~\ref{cal_table} are also applicable, in the first approximation, to individual group areas.
   Therefore, we construct the composite series of individual group areas using the corresponding $\rm{``b"}$ values from Table~\ref{cal_table}.

%============================PSI=====================
\subsection{Photometric Sunspot Index (PSI)} \label{section_psi}

In this section we present a daily Photometric Sunspot Index (PSI) series since 1874. PSI  \citep{1982SoPh...76..211H,1994SoPh..152..119B} is widely used in empirical irradiance reconstructions. PSI is a simple measure of reduction in solar output due to the presence of spots on the visible solar disc. 

Quantitatively, the suppression of the radiative output due to a single spot is defined as:
 \begin{equation}
  \Delta S_S = \frac{\mu A_S(C_S-1)(3\mu +2)}{2}.
\end{equation}
Here A$_S$ is the individual projected sunspot group area and $\mu$ is the cosine of the heliocentric angle. The quantity $(C_S-1)$ represents the residual intensity contrast of a sunspot with respect to the quiet photosphere. Following \citet{1992sers.conf..130B,1994SoPh..152..119B,1994SoPh..152..111F}, we calculate it as 
\begin{equation}
  C_S - 1 = 0.2231 + 0.0244 \cdot \log(A_S).
\label{eq3}
\end{equation}
The contributions from individual spots are summed up to derive the PSI as: 
\begin{equation}
 P_S  =\sum_{i=1}^{n} \left( \frac{\Delta S_S}{S_Q}\right)_i,
 \label{psi_eqn}
\end{equation}
where n is the total number of spots on the disc on a particular day. The result is expressed in units of $\mathrm{S}_Q$, the quite-Sun solar irradiance which is taken as 1361 $\rm{W/m^2}$ \citep{2011GeoRL..38.1706K}.
%----------------------------
\begin{figure*}[!h]
\centering
\includegraphics[width=0.85\textwidth]{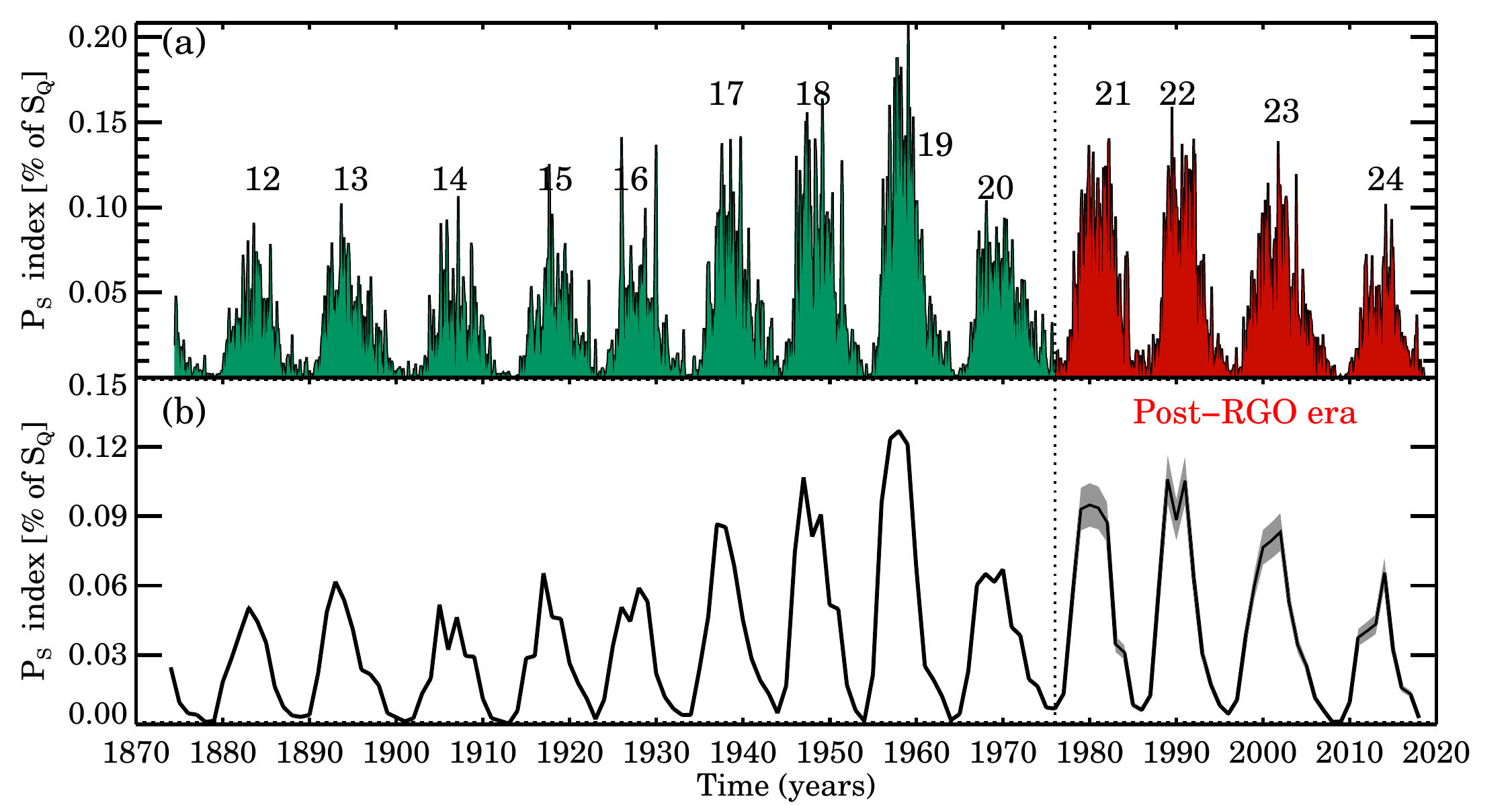}
\caption{Generated PSI series for monthly (top panel) and yearly (bottom panel) averaged data. Calculated error values (Grey shaded regions) are only shown for the yearly series. The dotted vertical line marks the year 1976, when RGO stopped its observing campaign.} 
\label{psi_series}
\end{figure*}
%------------------------------

\begin{figure*}[!h]
\centering
\includegraphics[width=0.85\textwidth]{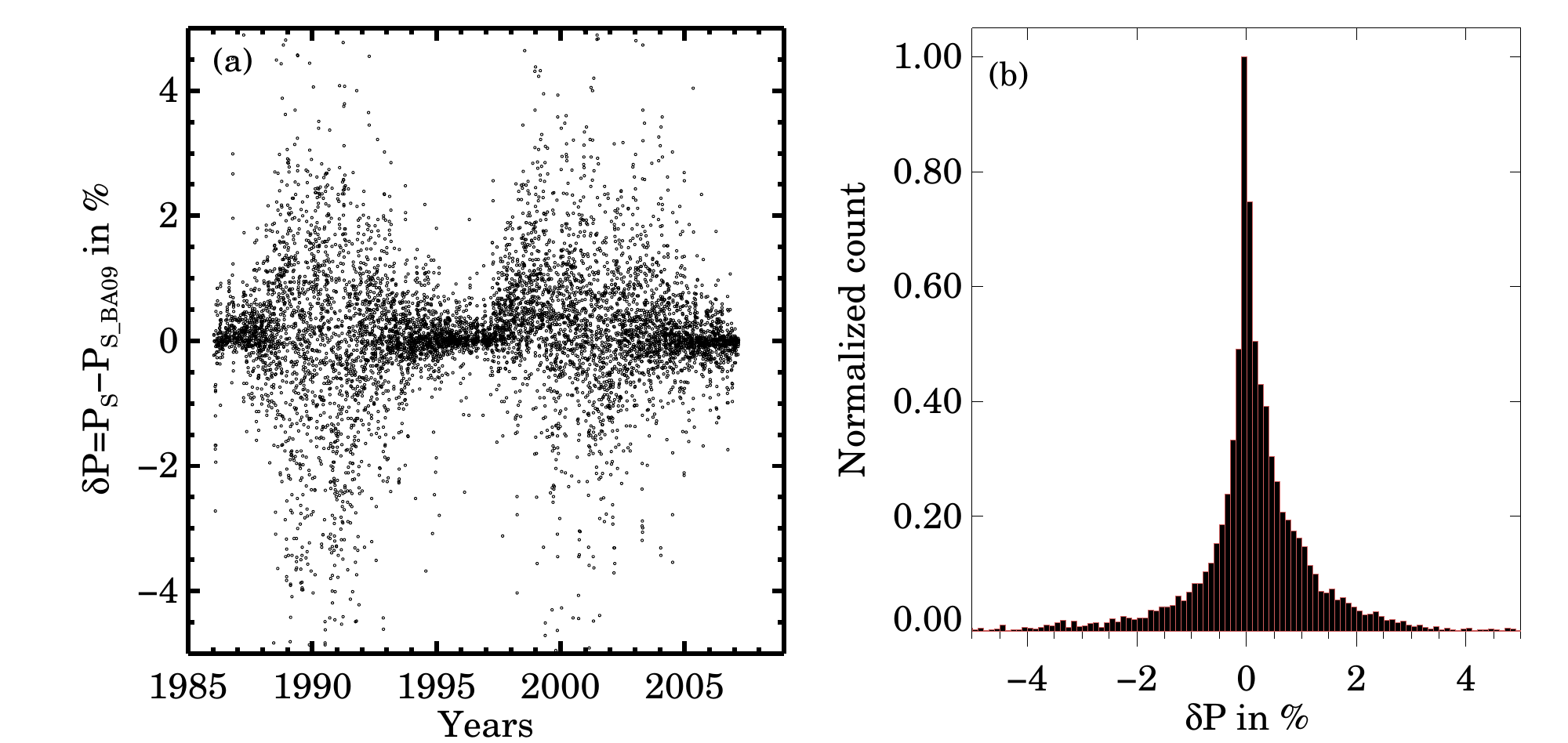}
\caption{a: Relative differences ($\delta$P in \%) of daily PSI values between this study ($\mathrm{P_S}$) and BA09 ($\mathrm{P_{S\_BA09}}$). b: a histogram of $\delta$P values.}
\label{psi_compare}
\end{figure*}

We calculate the daily PSI series by plugging in area values from our calibrated individual group area series into Eq.~\ref{psi_eqn}. The monthly and yearly values are plotted in Fig.~\ref{psi_series}a and b, respectively. Shaded regions in Panel~\ref{psi_series}b highlight the upper and lower limits of $\mathrm{P_S}$, which are generated using the two extreme limits of calibrated areas shown in Fig.~\ref{cal_area}. Next, we compare our PSI series ($\mathrm{P_S}$) with the PSI values from BA09 ($\mathrm{P_{S\_BA09}}$). We only perform it for the period of 1986-2008 where the SOON data were used by BA09 and the results are plotted in Fig.~\ref{psi_compare}. Looking at the plot, we conclude that the differences ($\delta\mathrm{P}$=$\rm{P_S}$--$\rm{P_{S\_BA09}}$) are small and are mostly below 1\%. Differences in the derived PSI values between the two series can be due to multiple reasons. Errors in sunspot area measurement is one such source and, by definition of PSI (Eq.~\ref{eq3}), errors in the measured spot positions (via $\mu$) also contribute to it. Hence, the true errors associated with individual PSI values are possibly slightly larger than our current estimate. 
In recent years, some studies (e.g., \citealp{2014SoPh..289.1517F}) claimed that missing small spots in sunspot area catalogues may introduce larger uncertainties in the derived PSI values due to the different contrast of small and big spots. Now, the PSI series of BA09 between 1986-2008 was constructed using the SOON catalogue which is known to have regularly missed smaller spots. A comparison of that series with our values (which includes smaller spots) shows small differences below 1\%. Thus, the errors in PSI introduced by the calibration of records that
miss small spots (such as SOON) seem to be low. Again, a further detailed study including individual `spot-to-spot' comparisons, is necessary to confirm this conclusion.

%---------------------------
%---------------------------

%+================Summary and Conclusion=============
\section{Summary and Conclusion} \label{section_summary}

A number of observatories around the globe carried out measurements of sunspot areas and positions over the last century.
RGO, the longest sunspot area database to date, started its campaign in 1874 and after continuing for a century, stopped it in 1976. Several other observatories from different parts of the world (e.g. Kodaikanal, Kislovodsk, Debrecen, Rome etc.), also carried out such observing programs during the 20th century. 
Sunspot area datasets are invaluable historical records of solar magnetic fields and are key to understanding the solar variability and its historical reconstructions. Hence, a long and consistent area series is expected to be of considerable use to the solar community. However, area measurements in each of these datasets are different from the others and hence, a merger is not a trivial task.

In this work, we have analysed and compared sunspot group areas from a total of nine observatories (RGO, Kislovodsk, Pulkovo, Debrecen, Kodaikanal, SOON, Rome, Catania, Yunnan). It turned out that data from only four observatories (RGO, Kislovodsk, Pulkovo, Debrecen) are sufficient to produce cross-calibrated, up-to-date (1874-2019) catalogues of daily total and individual group areas. The remaining gaps (776 days in total) could not be filled with data from the other archives as the missing days lie either before 1922 or after 2016 and none of the other archives covers these periods.
For completeness, we still list the derived scaling coefficients for all the data sets in Table~\ref{cal_table}, as future studies might perhaps find this useful.
As in the earlier studies, we found that areas from Kislovodsk and Pulkovo observatories are in good agreement with RGO, while also having a very good temporal coverage.
This is a significant advantage over the previous similar studies in which composite of total sunspot area time series were generated using SOON areas,
in which sunspot areas are 50\% smaller compared to RGO measurements. Along with that, SOON does not have any direct overlap with RGO whereas both Kislovodsk and Pulkovo, used in our series, have long overlaps.
The choice of these observatories in constructing this catalogue is further justified by our analysis of the variation of the calibration factors with time. We find that our chosen observatories (RGO, Kislovodsk, Pulkovo, Debrecen) are significantly more stable than the other observatories (SOON, Kodaikanal, Yunnan, Rome).
In fact, just RGO and Kislovodsk together cover 94\% of the observing days between 1874 and 2019. Overall, the use of Kislovodsk (and Pulkovo) helped us to reduce the uncertainties in the generated catalogue. The remaining gaps are filled with areas from Debrecen which also has similar area measurements as RGO (with the calibration factor $\rm{``b"}$ being 1.06). 
Thus, our entire catalogue is made out of observations which are either directly from RGO or very similar to RGO in their properties.
This increases the quality as well as the reliability of our catalogue. In this paper, we also compared data from Kodaikanal, SOON, Rome and Catania. Our results show that, although some of these data sets cover long periods (e.g. SOON and Kodaikanal), their area measurements are rather significantly different from  RGO and, more importantly, display considerable scatter and/or trends compared with the other observatories.

We have compared our area values to the earlier version of the composite by BA09. 
In particular, by using the Kislovodsk data  we have accounted for a systematic offset between roughly 1977 and 1985 which was present in BA09's series. This offset, already noted by BA09 earlier, was due to the use of old Russian data in their series.
Compared to the sunspot number, the scatter in our area values is smaller than in BA09. We emphasize, however, the need for an in-depth, `spot-to-spot' calibration study to address some complicated individual cases. In addition to the corrected areas, we also provide a calibrated projected area series and a preliminary series of individual group areas. 
Furthermore, by using our calibrated area catalogue we have calculated the daily PSI, which is often used in irradiance reconstructions. 
Compared to the earlier PSI record from BA09, we found that the effect of `missed small spots' (e.g. due to their usage of the SOON data) onto the calculated PSI, is not significant.

To take this work further, we plan to add more data sets in the future. Sunspot data from four Chinese stations: Qingdao Observing Station, Purple Mountain Astronomical Observatory, Yunan Astronomical Observatory and Chinese Solar-Geophysical data have recently been digitized (including the parameter extractions) \citep{2019arXiv190412316L}. These sets of data cover almost 90 years (1925--2015) and will be a great source to further improve the catalogue. The other followup work planned in this context, is to perform a `spot-to-spot' calibration between observatories. This will basically be a detailed comparison (such as the shape and size) of every sunspots that has been simultaneously recorded by multiple observatories. Such an analysis will help us to better understand the dependency of measurement errors on a particular spot size. Also, it could also provide insights on quantifying the time variation effects within an observatory.

All catalogues produced here are available online with this publication and at \url{http://www2.mps.mpg.de/projects/sun-climate/data.html}.

%==================Ack======================
\section{Acknowledgment}

We thank the anonymous reviewer for the encouraging comments and helpful suggestions.
We also thank the teams of the archives used in this study for all the work they had invested into obtaining and making these data available to the community.

 \bibliographystyle{aa}
% \bibliography{references_MPS}

\end{document}